\documentclass[sigconf,natbib=true,anonymous=false,review=false]{acmart}

\usepackage{algorithm}
\usepackage{algorithmic}
\usepackage{multirow}
\usepackage{enumitem}
\usepackage{caption}
\usepackage{booktabs}
\usepackage{amsmath,amsfonts}
\usepackage{textcomp}
\usepackage{subcaption}
\usepackage{footnote}
\usepackage{graphicx}
\usepackage{multirow}
\usepackage{graphicx}
\usepackage{enumitem}
  
\allowdisplaybreaks[4]
\usepackage{hyperref}

\theoremstyle{definition}

\newcommand\M{AR-Trip}

\AtBeginDocument{%
  \providecommand\BibTeX{{%
    \normalfont B\kern-0.5em{\scshape i\kern-0.25em b}\kern-0.8em\TeX}}}

\copyrightyear{2024}
\acmYear{2024}
\setcopyright{acmlicensed}
\acmConference[SIGIR '24] {Proceedings of the 47th International ACM SIGIR Conference on Research and Development in Information Retrieval}{July 14--18, 2024}{Washington, DC, USA.}
\acmBooktitle{Proceedings of the 47th International ACM SIGIR Conference on Research and Development in Information Retrieval (SIGIR '24), July 14--18, 2024, Washington, DC, USA}
\acmISBN{979-8-4007-0431-4/24/07}
\acmDOI{10.1145/3626772.3657970}

\begin{document}

\title{Analyzing and Mitigating Repetitions in Trip Recommendation}

\author{Wenzheng Shu}
\authornote{Contributed equally to this research.}
\affiliation{
  \institution{University of Electronic Science and Technology of China}
  \city{Chengdu}
  \country{China}
}
\email{shuwenzheng926@gmail.com	}

\author{Kangqi Xu}
\authornotemark[1]
\affiliation{
  \institution{University of Electronic Science and Technology of China}
  \city{Chengdu}
  \country{China}
}
\email{supanity@outlook.com}

\author{Wenxin Tai}

\affiliation{
  \institution{University of Electronic Science and Technology of China}
  \city{Chengdu}
  \country{China}
}
\email{wxtai@std.uestc.edu.cn}

\author{Ting Zhong}

\affiliation{
  \institution{University of Electronic Science and Technology of China}
  \city{Chengdu}
  \country{China}
}
\email{zhongting@uestc.edu.cn}

\author{Yong Wang}

\affiliation{
  \institution{Hong Kong University of Science and Technology}
  \city{Hong Kong}
  \country{China}
}
\email{wangyongjoy@ust.hk}

\author{Fan Zhou}
\authornote{Corresponding author.}
\affiliation{
  \institution{University of Electronic Science and Technology of China}
  \institution{Intelligent Terminal Key Laboratory of Sichuan Province}
  \country{China}
}
\email{fan.zhou@uestc.edu.cn}

\renewcommand{\shortauthors}{Wenzheng Shu et al.}

\begin{abstract}
   Trip recommendation has emerged as a highly sought-after service over the past decade. Although current studies significantly understand human intention consistency, they struggle with undesired repetitive outcomes that need resolution. We make two pivotal discoveries using statistical analyses and experimental designs: (1) The occurrence of repetitions is intricately linked to the models and decoding strategies. (2) During training and decoding, adding perturbations to logits can reduce repetition. Motivated by these observations, we introduce \textbf{\M} (\textbf{A}nti \textbf{R}epetition for \textbf{Trip} Recommendation), which incorporates a cycle-aware predictor comprising three mechanisms to avoid duplicate Points-of-Interest (POIs) and demonstrates their effectiveness in alleviating repetition. Experiments on four public datasets illustrate that \M~ successfully mitigates repetition issues while enhancing precision.
\end{abstract}



\maketitle

\section{Introduction}
\label{subsec:intro}
The surge in location-based social networks (LBSNs), like Foursquare and Flickr, has served as a platform for generating a substantial volume of media data. The wealth of geospatial data holds immense promise across various areas, spanning next POI prediction~\cite{liu2016predicting,yang2022getnext}, duration estimation~\cite{lim2015personalized,xu2022mtlm}, human social interaction~\cite{wu2019graph}, route planning~\cite{xu2015efficient,jain2021neuromlr} and trip recommendation~\cite{zhou2021contrastive,gao2022self,zhao2023gc}. Unlike the minimization strategy of route planning, trip recommendation shows considerable promise. It prioritizes addressing user intentions and generating results that reflect user preferences in practical scenarios~\cite{zhou2019context}.

\noindent \textbf{Related work.}
\label{subsec:rela}
Previous studies have mainly focused on planning-based methods tailored for the Orienteering Problem (OP). For instance, Popularity~\cite{chen2016learning} considers only POI popularity, PersTour~\cite{lim2015personalized} includes user preferences, and Markov~\cite{chen2016learning} uses POI transition relations. Subsequent approaches go beyond treating recommendation purely as an optimization problem. POIRank~\cite{chen2016learning} integrates multiple attributes using RankSVM, while RankMarkov~\cite{chen2016learning} merges Markov and POIRank. Additionally, C-ILP~\cite{he2019joint} introduces context-aware POI embedding through linear programming. However, these planning-based solutions fail to fully capture the complexity of human mobility, prompting the adoption of learning-based techniques to address these challenges. CTLTR~\cite{zhou2021contrastive} and SelfTrip~\cite{gao2022self} use RNNs~\cite{elman1990finding}, while DeepTrip~\cite{gao2021adversarial} incorporates GANs~\cite{gao2021adversarial}. Transformer-based models~\cite{vaswani2017attention} like Bert-Trip~\cite{kuo2023bert} treat trip recommendation as a sentence completion task to offer personalized itineraries aligning with tourists' interests and constraints~\cite{lim2019tour}.

\noindent \textbf{Challenge.}
\label{subsec:chall}
The techniques mentioned above still unavoidably encounter \textbf{t}rip \textbf{r}epetition \textbf{p}roblems (TRPs) -- the phenomenon of generating duplicate recommendations at the POI level. Indeed, prior studies in planning-based methods~\cite{chen2016learning,anagnostopoulos2017tour,taylor2018travel} have attempted to explore such issues both through statistical and practical investigations, including integer linear program (ILP)~\cite{chen2016learning}, branch-and-bound algorithm~\cite{taylor2018travel} and budget-constraint pruning optimization~\cite{anagnostopoulos2017tour}. Although heuristic-based researches can alleviate (or even eliminate) repetition, they compromise on the accuracy of predictions. Correspondingly, despite the high accuracy of learning-based methods, they struggle to explain these repetitive patterns and lack effective solutions, hindering the usability of generated itineraries.

\noindent \textbf{Present work.}
\label{subsec:pres}
Inspired by the recent flourishing development of large language models (LLMs)\cite{radford2018improving}, we recognize the potential similarity between our tasks and text-driven ones regarding their ability to understand and mimic human behaviors\cite{holtzman2019curious}. Despite numerous relevant studies~\cite{fu2021theoretical,foster2007avoiding} addressing language degeneration, their methods do not apply to our tasks. On the one hand, language degradation mainly entails the repetition of sentences (or sub-sentences), enhancing the available contextual information~\cite{xu2022learning}. In contrast, the trip recommendation is a challenging NP-hard task~\cite{zhou2021contrastive} relying on limited POI queries and historical knowledge as its primary inputs. Moreover, language degradation primarily focuses on the self-looping phenomenon among sentences, which has clear boundaries regarding sentence repetition~\cite{xu2022learning}. In trip recommendation, self-looping is just one form of repetition. However, due to the constraints imposed by human mobility patterns, repetitive generation within trajectories is prohibited~\cite{chen2016learning,gao2018trajectory}.
After comprehending the difference above, we empirically validate the potential causes of trajectory repetition. Meanwhile, we present a pioneering cycle-aware framework that applies three mechanisms to reconstruct the logits in both the training and decoding phases. Our main contributions are summarized as threefold:

\begin{itemize}[leftmargin=*]
    \item We analyze two potential causes of TRPs in the domain of trip recommendation for the first time.
    \item We introduce a fresh perspective to tackle this issue of repetitions and propose three cycle-aware methods to mitigate it, taking both the training and decoding phases into account.
    \item We introduce a novel metric to assess repetition within trip recommendations, and the results demonstrate the effectiveness of our approach in significantly reducing repetition.
\end{itemize}

\section{Preliminary}
\subsection{Problem Definition}
\label{subsec:problem}
We introduce symbolic representations of trajectories for subsequent chapters. At the $i$-th position, $p_i \in \mathcal{P}$ and $t_i$ represent the actual POI and time, while $\hat{p}_i$ and $\hat{t}_i$ represent their generated counterparts, where $\mathcal{P}$ denotes the set of POIs. A route of length $m$ can be denoted as $R:[R_p,R_t] \Rightarrow[\{p_i\}_{i=1}^m,\{t_i\}_{i=1}^m] \in \mathcal{R}$, where $\mathcal{R}$ indicates the set of historical trajectories. The query $Q:\{p_s, t_s, p_e, t_e, n\}$ includes start and end points with timestamps and the desired length $n$ of the trajectory. $\hat{T}:\{\hat{p}_i\}_{i=1}^n$ represents the generated results. The task can be formulated as follows: the traveler provides $Q$ as input to the model, which utilizes latent travel intentions from $\mathcal{R}$ to generate the predicted path $\hat{T}$.

\subsection{Overview of Repetitive Patterns}
\label{subsec:Repetition}

\begin{figure}[t]
   \centering
   \begin{subfigure}[b]{0.23\textwidth}
      \centering
      \includegraphics[width=\textwidth]{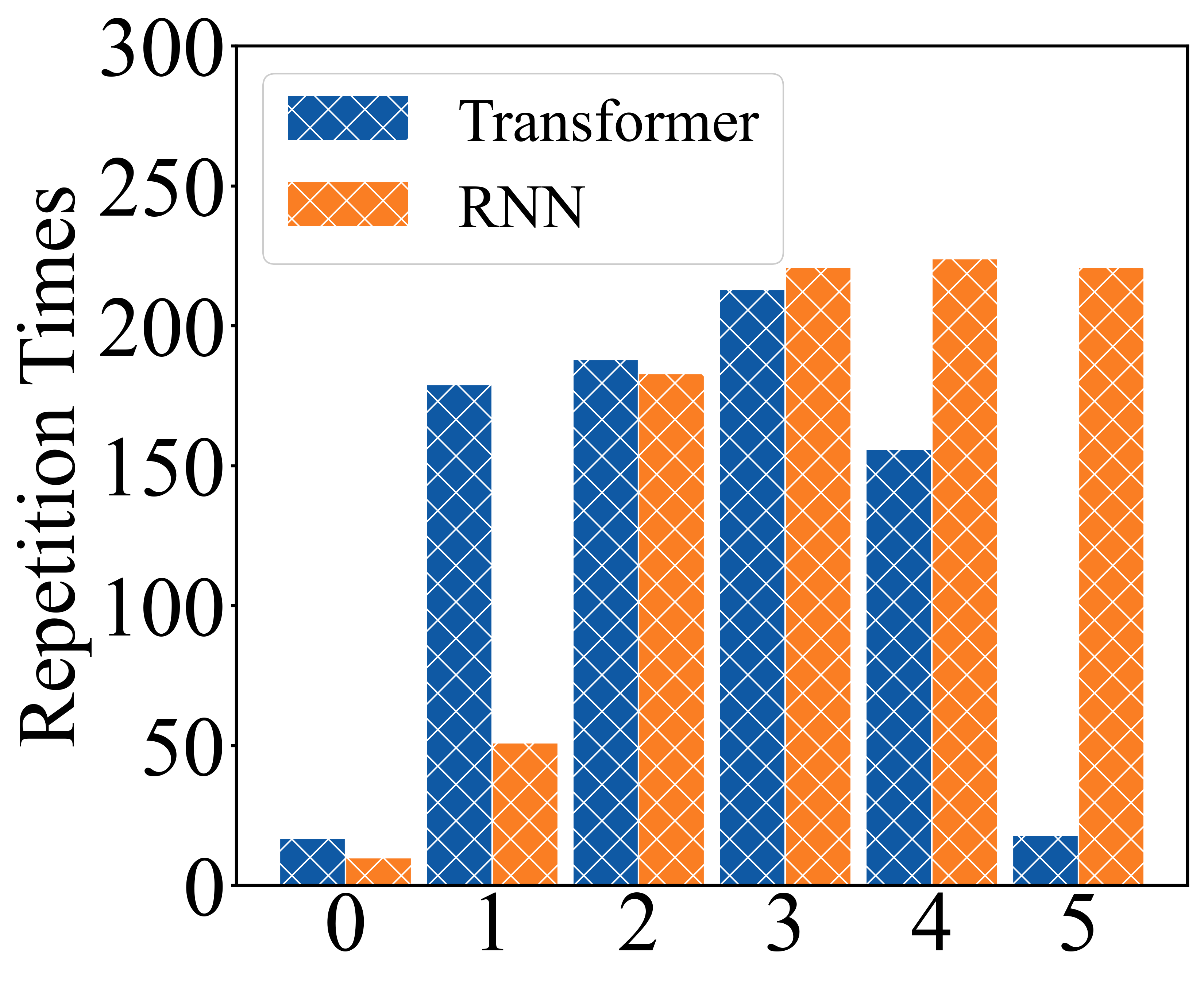}
      \caption{Repetition(length=6)}
      \label{fig:pics_1}
   \end{subfigure}
   \hfill
   \begin{subfigure}[b]{0.23\textwidth}
      \centering
      \includegraphics[width=\textwidth]{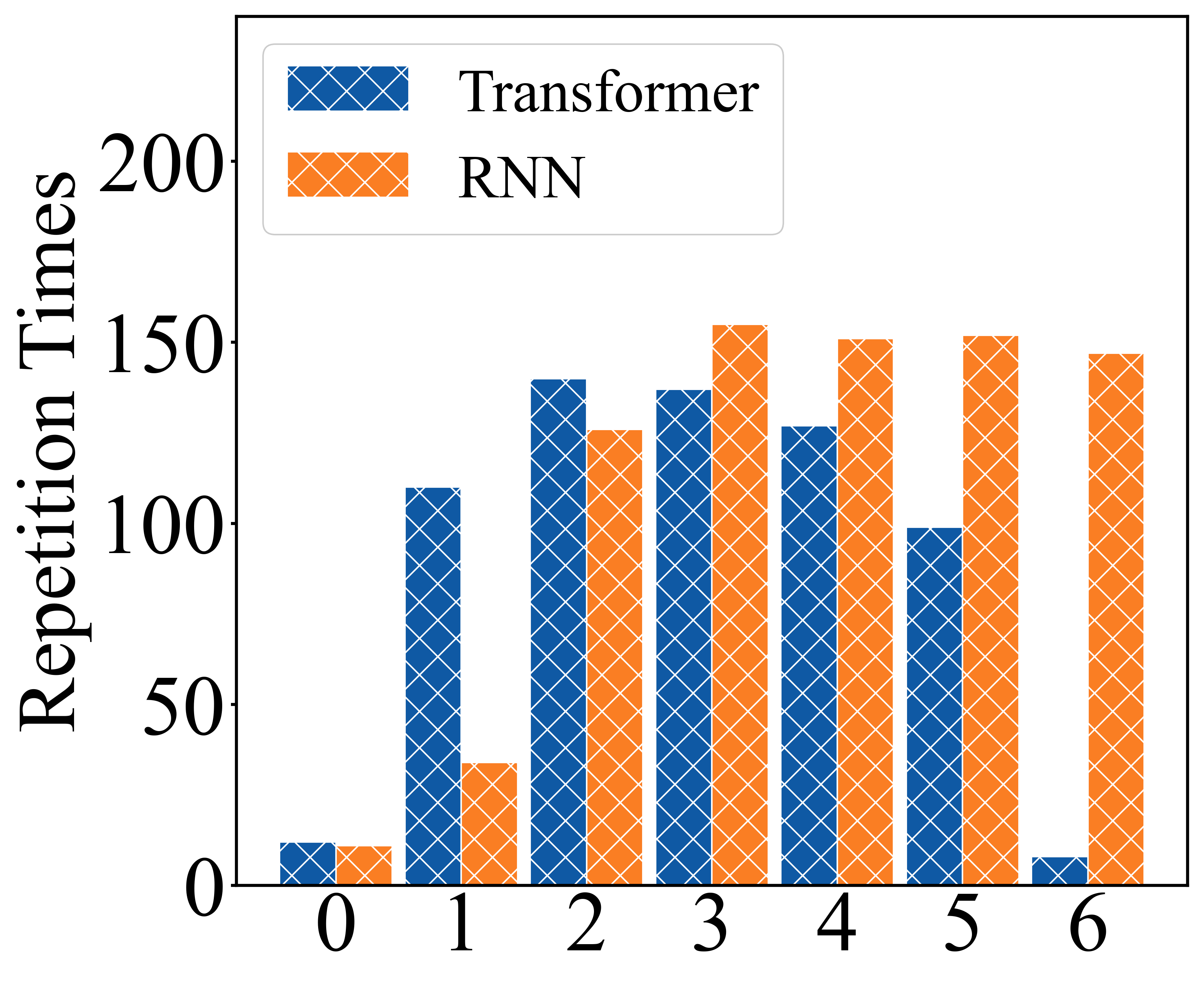}
      \caption{Repetition(length=7)}
      \label{fig:pics_2}
   \end{subfigure}
   \caption{Model selection: we showcase trajectories of lengths six (Figure \ref{fig:pics_1}) and seven (Figure \ref{fig:pics_2}), which are data-rich and long enough for analysis.}
   \label{fig:pics_1&2}
   \vspace{-0.5cm}
\end{figure}

\noindent \textbf{Training Models.}
\label{subsec:training}
Previous studies~\cite{holtzman2019curious,fu2021theoretical} have underscored the connection between repetition and the intrinsic framework of the model architecture~\cite{holtzman2019curious}. To further investigate the correlation between models and repetitions in the trip recommendation, we employ standard RNN~\cite{elman1990finding} and Transformer~\cite{vaswani2017attention} commonly utilized in this domain to assess how frequently POIs repeat relative to their positions in previous and subsequent visits. Figure~\ref{fig:pics_1&2} indicates that compared to the RNN model, the Transformer exhibits a significantly lower repetition frequency, especially in the latter half of the trajectory. This discrepancy primarily arises from the differences in encoding characteristics and information processing between RNNs and Transformers~\cite{kuo2023bert}. Due to its structural limitations and the implicit representation of queries, RNNs are more susceptible to the influence of previously generated sequences~\cite{gao2022self}, leading to increased repetitive content in the latter segments of the sequence. In contrast, Transformers can directly utilize endpoint information as explicit conditions~\cite{ho2022poibert}, thereby ensuring the accuracy and coherence of trajectory predictions at both the beginning and end.

\noindent \textbf{Decoding Strategies.}
\label{subsec:decoding}
Expanding on the insights from~\cite{fu2021theoretical}, we investigated the correlation between POI transition probabilities and decoding for generation. Considering each POI at position $i$ is associated with a sparse transition probability matrix $M_i$, we quantify this sparsity using $\xi = \frac{1}{k^2}\sum_{i=1}^{k}\sum_{j=1}^{k} \mathbb{I}$, where $k=|\mathcal{P}|$ indicates the number of POIs and $\mathbb{I}$ equals to 1 if it's true. In our task, learning-based methods~\cite{gao2021adversarial,gao2022self,zhou2021contrastive} incorporate long-term dependencies, necessitating the addition of i.i.d noise $N_i$ to each $M_i$, resulting in $M_i' = M_i + N_i$. To evaluate the likelihood of repetition along the trajectory, we introduce the \textbf{P}robability \textbf{M}atrix of \textbf{R}epetition (PMR) as $\sum_{j=1}^{\infty }tr[\prod_{i=1}^{2j}M_i'/(k\xi)^{j}]$, where $\xi$ measures the proportion of non-zero probabilities in $M'$.
The finding suggests that reducing recommendation repetition hinges on decreasing the maximum eigenvalue probability of $\prod_{i=1}^{2j}M_i'$, thereby reducing the variance of probabilities across $M'$. Traditional learning-based predictors mainly employ the Greedy sampling strategy~\cite{gao2021adversarial,zhou2021contrastive,gao2022self}, aiming to maximize the likelihood~\cite{welleck2020neural} scores. However, studies~\cite{holtzman2019curious,fu2021theoretical} have shown that such deterministic encoding methods lead to repetitive patterns. Specifically, the Greedy tends to yield a minimum value of $\xi = 1/k$, which correlates with a higher PMR score~\cite{fu2021theoretical}.
Recent studies have introduced stochastic sampling methods to address this issue, including Top-k~\cite{fan2018hierarchical} and Top-p~\cite{holtzman2019curious}. These techniques utilize probability distributions to select optimally and truncate candidate sets, minimizing repetition and lowering the PMR~\cite{holtzman2019curious}.

\section{Methodology}
\subsection{Basic Architecture}
\begin{figure}[h]
    \centering
    \includegraphics[width=1.0\linewidth]{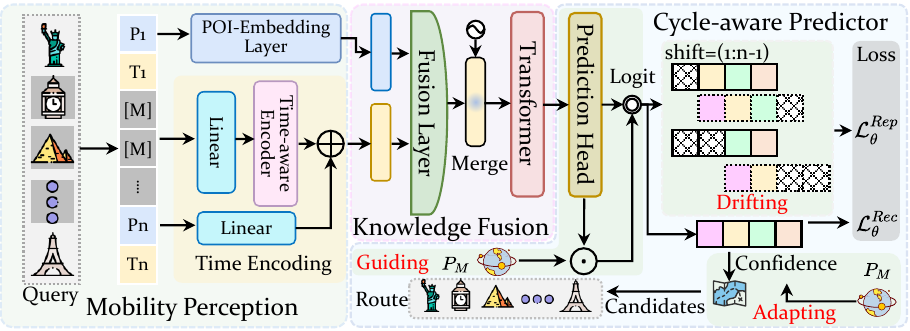}
    \caption{An overview of the proposed~\M~framework.}
    \label{fig:pics_3}
\end{figure}

Figure \ref{fig:pics_3} illustrates the basic framework of \M\footnote{The source code is available at \url{https://github.com/Joysmith99/AR-Trip}.}, which consists of three main components: mobility perception, knowledge fusion, and cycle-aware predictor. The preceding two components are built upon the encoding framework, resulting in Transformer outputs $Z_{out}$, which are then passed to the cycle-aware predictor.

\subsection{Cycle-aware Predictor}
Motivated by~\cite{fu2021theoretical} and preceding analysis in \textsection \ref{subsec:Repetition}, we acknowledge the potential correlation between $M'$ and logits. Therefore, our approach focuses on granting logits the ability for cyclical perception during training and decoding stages~\cite{xu2022learning}. Thus, we employ three methods: (1) The guiding matrix to incorporate prior knowledge and reshape logits; (2) The drifting optimization to steer logits towards unlikelihood tasks~\cite{welleck2020neural}; (3) The adapting candidates for constrained stochastic sampling.

\noindent \textbf{Guiding.}
Upon passing $Z_{out}$ through the prediction head, we acquire the POI logits $\mathcal{H}:\{h_i\}_{i=1}^m \in \mathbb{R}^{m \times \lvert \mathcal{P} \rvert}$. To explicitly guide $\mathcal{H}$, we then apply a position-based guidance matrix $P_M \in \mathbb{R}^{\lvert \mathcal{P} \rvert \times m_{max}}$ from $\mathcal{R}$, with $m_{max}$ representing the maximum route length in $\mathcal{R}$. $P_M$ is determined by the ratio ${f_{ij}}/{f_i} (1 \leq i \leq \lvert \mathcal{P} \rvert, 1 \leq j \leq m_{max})$, where $f_{ij}$ indicates the frequency of the $i$-th POI at position $j$, and $f_i$ signifies the total occurrences of the $i$-th POI in $R$. Subsequently, the guided logits $\hat{\mathcal{H}}:\{\hat{h}_i\}_{i=1}^m \in \mathbb{R}^{m \times \lvert \mathcal{P} \rvert}$ are obtained via $P_M$:
\begin{align}
\label{eq:4}
    \hat{\mathcal{H}} = \mathcal{H} + \mathcal{H} \odot P_{M}^T\{:m\},
\end{align}
where $\odot$ denotes the element-wise product, and $P_{M}^T\{:m\} \in \mathbb{R}^{m \times \lvert \mathcal{P} \rvert}$ signifies the alignment of the route to $m$ along with transposition.

\noindent \textbf{Drifting.}
After acquiring $\hat{\mathcal{H}}$, we integrate it into drift penalty training to better distinguish location-based differences among POIs. Giving $\hat{h}_i \in \mathcal{\hat{H}}$, the optimization procedure is delineated as follows:
\begin{align}
\label{eq:4}
   \mathcal{L}_\theta^{Rep} = -\sum_{\omega=1}^{m-1}\sum_{i=1}^{m-\omega}\log(1-Pr(\hat{h}_i, \hat{h}_{i+\omega})),
\end{align}
where $\omega$ is the shift parameter and $Pr(\cdot)$ represents the probability from normalized cosine similarity between $\hat{h}_i$ and $\hat{h}_{i+\omega}$. Ultimately, we adopt a multi-task learning strategy and Adam optimizer for model optimization, drawing on prior research~\cite{kuo2023bert}:
\begin{align}
\label{eq:4}
    \mathcal{L}_{\theta} = \mathcal{L}_{\theta}^{Rec} + \alpha \cdot \mathcal{L}_{\theta}^{Rep},
\end{align}
where $\mathcal{L}_{\theta}^{Rec}$ is equivalent to the standard cross-entropy (CE) loss applied in the mapping of POIs, with $\alpha$ serving as a hyperparameter.

\noindent \textbf{Adapting.}
Section \textsection \ref{subsec:pres} highlights the particularity of trip recommendation, and it is not reasonable to directly implement stochastic sampling through the fixed threshold. Therefore, we introduce a stochastic encoding strategy utilizing confidence score $\mathcal{C}$, which mitigates repetition but is also constrained by $\mathcal{C}$ itself. Unlike the temperature in prior work~\cite{holtzman2019curious}, $\mathcal{C}$ is derived from the $P_M$. We pioneer an advanced stochastic decoding method by combining nuclear sampling with $\mathcal{C}$, which can be formulated as $\{f_j^0\}/|\mathcal P|(1\leq j\leq m_{max})$. Here, $\{f_{j}^0\}$ signifies the count of POIs with no occurrence in the $j$-th position at $P_M$. In other words, when a specific location presents several possible candidates, $\mathcal{C}$ lowers confidence to smooth out the logits, ensuring diversity and reducing repetition without sacrificing accuracy (see results in \textsection \ref{subsec:performance}).

\section{Experiments}
We now conduct experiments on four real-world datasets to assess the results.

\begin{table*}[h]
    \centering
    \setlength{\tabcolsep}{4pt}
    \renewcommand{\arraystretch}{0.5}
    \caption{Evaluation of four datasets. The best performance is emphasized in bold, while the second-best results are underlined. The hyphen (---) signifies the utilization of statistical methods to prevent repetition. It is important to note that C-ILP has the potential to generate longer trips, resulting in a higher $F_1$ value, distinguished by an asterisk (*). In this case, the $PairsF_1$ cannot be calculated and is represented by a cross (X) for distinction.}

    \label{tab:experiments} 
    \begin{tabular}{@{}c|ccc|ccc|ccc|ccc@{}}  
        \toprule
        \textbf{Dataset}&\multicolumn{3}{c|}{\textbf{Edinburgh}}&\multicolumn{3}{c|}{\textbf{Glasgow}}&\multicolumn{3}{c|}{\textbf{Osaka}} &\multicolumn{3}{c}{\textbf{Toronto}}\cr 
        \midrule  
        \textbf{Metric}&\multicolumn{1}{c}{$F_1$}&\multicolumn{1}{c}{$PairsF_1$}&\multicolumn{1}{c|}{$REP$}&\multicolumn{1}{c}{$F_1$}&\multicolumn{1}{c}{$PairsF_1$}&\multicolumn{1}{c|}{$REP$}
        &\multicolumn{1}{c}{$F_1$}&\multicolumn{1}{c}{$PairsF_1$}&\multicolumn{1}{c|}{$REP$}
        &\multicolumn{1}{c}{$F_1$}&\multicolumn{1}{c}{$PairsF_1$}&\multicolumn{1}{c}{$REP$}\cr
        \midrule
        Popularity&0.701&0.436&---&0.744&0.505&---&0.663&0.365&---&0.678&0.384&---\cr
        PersTour&0.656&0.417&---&0.801&0.643&---&0.686&0.468&---&0.720&0.504&---\cr
        POIRank&0.700&0.431&---&0.769&0.550&---&0.745&0.511&---&0.753&0.517&---\cr
        C-ILP&0.743*&X&---&0.835*&X&---&0.752*&X&---&0.810*&X&---\cr
        Markov&0.645&0.417&0.092&0.725&0.493&\underline{0.030}&0.697&0.445&\underline{0.033}&0.669&0.407&0.042\cr
        RankMarkov&0.662&0.448&0.095&0.753&0.545&0.035&0.715&0.486&0.037&0.722&0.511&\underline{0.035}\cr
        DeepTrip&0.638&0.528&0.238&0.714&0.597&0.165&0.746&0.634&0.145&0.680&0.558&0.183\cr
        CTLTR&0.728&0.681&0.211&0.838&0.763&0.087&0.802&0.719&0.112&0.806&0.748&0.139\cr
        SelfTrip&0.728&\underline{0.688}&0.199&0.834&\underline{0.773}&0.116&0.840&\underline{0.774}&0.117&0.790&\underline{0.751}&0.134\cr
        Bert-Trip&\underline{0.814}&0.672&\underline{0.066}&\underline{0.849}&0.740&0.043&\underline{0.854}&0.740&0.052&\underline{0.859}&0.733&\underline{0.035}\cr
        \midrule
        \textbf{\M}&\textbf{0.868}&\textbf{0.808}&\textbf{0.049}&\textbf{0.881}&\textbf{0.820}&\textbf{0.027}&\textbf{0.871}&\textbf{0.828}&\textbf{0.016}&\textbf{0.892}&\textbf{0.839}&\textbf{0.024}\cr
        \bottomrule
    \end{tabular}
\end{table*}

\begin{table}[ht]
    \centering
    \setlength{\tabcolsep}{4pt}
    \renewcommand{\arraystretch}{0.8}
    \caption{Performance of different elements setting. Note that: (+) add  (a) Adapting candidates; (g) Guiding matrix; (d) Drifting optimization.}
    \label{tab:ablation}
    \resizebox{\linewidth}{!}{
    \begin{tabular}{l|rrr|l|rrr}
    \toprule
        \textbf{elements}&$F_1$&$PF_1$&$REP$&\textbf{elements}&$F_1$ & $PF_1$ & $REP$\\ 
        \midrule
        +agd (Ours)&0.881&0.820&0.027&+null (Base)&0.850&0.799&0.096\\
        +ag&0.884&0.849&0.048&+a&0.886&0.852&0.062\\
        +ad&0.879&0.828&0.035&+d&0.851&0.767&0.050 \\
        +gd&0.848&0.761&0.051&+g&0.852&0.793&0.093 \\
    \bottomrule
    \end{tabular}
    }
    \vspace{-0.5cm}
\end{table}

\subsection{Experimental Setup}
\label{subsec:setup}
\noindent \textbf{Datasets.} We utilize Flickr \cite{lim2015personalized} datasets from four cities: Edinburgh, Glasgow, Osaka, and Toronto, following previous works~\cite{zhou2021contrastive}. In this dataset, check-ins were obtained from geo-tagged YFCC100M Flickr photos.
To ensure fairness, we conduct five repeated experiments for all baselines.

\noindent \textbf{Baselines.} We conduct a comparative analysis between our proposed framework and the following state-of-the-art methods, which encompass: (1) Traditional methods: Popularity~\cite{chen2016learning}, PersTour~\cite{lim2015personalized}, POIRank~\cite{chen2016learning}, C-ILP~\cite{he2019joint}, Markov and RankMarkov~\cite{chen2016learning}. (2) learning-based methods: DeepTrip~\cite{gao2021adversarial}, CTLTR~\cite{zhou2021contrastive}, SelfTrip~\cite{gao2022self} and Bert-Trip~\cite{kuo2023bert}. 

\noindent \textbf{Metrics.} We compare {\M} with other baselines using $F_1$ and $PairsF_1$ (both the values range between 0 and 1), following prior studies~\cite{lim2015personalized,chen2016learning,he2019joint}. In addition, to measure the repetitive problems, we devise an extra indicator $REP$ to calculate the ratio of non-repeating POIs to the total length of the trajectory: 
\begin{align}
    REP = \frac{1}{n\lvert \mathcal{R} \rvert} \sum_{R \in \mathcal{R}} (n-\mathcal{U}(R)),
\end{align}
 where $\mathcal{U}(R)$ signifies transforming the trajectory into a set of unique POIs. It is worth mentioning that $REP$ maintains consistency with the range of $F_1$ and $PairsF_1$. A higher $F_1$ and $PairsF_1$, or a lower $REP$, indicates better performance.

\noindent \textbf{Parameter Settings.}
For experimental settings, the embedding size was fixed at 32, covering POI, time, and position embeddings. The learning rate is set to 1e-3, and training spanned 50 epochs across four datasets, optimizing with $\alpha$ at 1.0 and the decoding threshold at 0.8. Our model utilized the PyTorch library on a GeForce RTX 3060 GPU with 12GB.

\subsection{Performance Comparisons}
\label{subsec:performance}
\noindent \textbf{Experimental Results.}
Table~\ref{tab:experiments} compares the prediction results between the baselines and our method. It is evident that~\M~ outperforms all the baseline methods across all metrics. While the statistical methods often exhibit very low (or nonexistent) $REP$ scores, they cannot ensure accuracy regarding $F_1$ and $PairsF_1$. On the other hand, learning-based methods achieve higher accuracy but introduce severe repetition issues. While Bert-Trip can balance $F_1$ and $REP$, it doesn't achieve the top $PairsF_1$ score among others. We realize that it stems from the calculation of $PairsF_1$ itself. For example, when compared to target sequence $\{4, 2, 8, 5\}$, a conventional decoder might produce $\{4, 2, 2, 5\}$. Despite repetition, it can achieve a score of 0.833 in $PairsF_1$. Even if a cycle-aware method might produce $\{4, 8, 2, 5\}$ with 100\% accurate $F_1$, it only attains a $PairsF_1$ score of 0.5, indicating that $PairsF_1$ overlooks the impact of repeated elements. Correspondingly, \M~effectively maintains accuracy while preventing repetition phenomena.

\noindent \textbf{Ablation Studies.}
\label{ablation}
To evaluate the effectiveness of the elements in our framework, we conduct relative experiments to compare the performance of our model in Glasgow, as shown in Table~\ref{tab:ablation}. From it, we can observe that: (1) The adapting and drifting elements mentioned above effectively mitigate the repetition issue, and they also show improvements in $F_1$ compared to the Base model. (2) Guiding-based modules become effective when working with adapting modules. This is due to the Transformer's one-shot output and unique attention mechanism~\cite{vaswani2017attention}, unlike traditional RNNs, securing long-term dependencies in recommendations. Essentially, combining guidance with Greedy in the Transformer model doesn't significantly alter outcomes compared to using Greedy alone, possibly because deterministic encoding maintains the original distribution $\mathcal{H}=\hat{\mathcal{H}}-\mathcal{H} \odot P_M^T$. (3) The experiments effectively demonstrate that our approach significantly reduces repetition while maintaining the accuracy of trajectory recommendations. 

\begin{figure}[ht]
   \centering
   \begin{subfigure}[b]{0.23\textwidth}
      \centering
      \includegraphics[width=\textwidth]{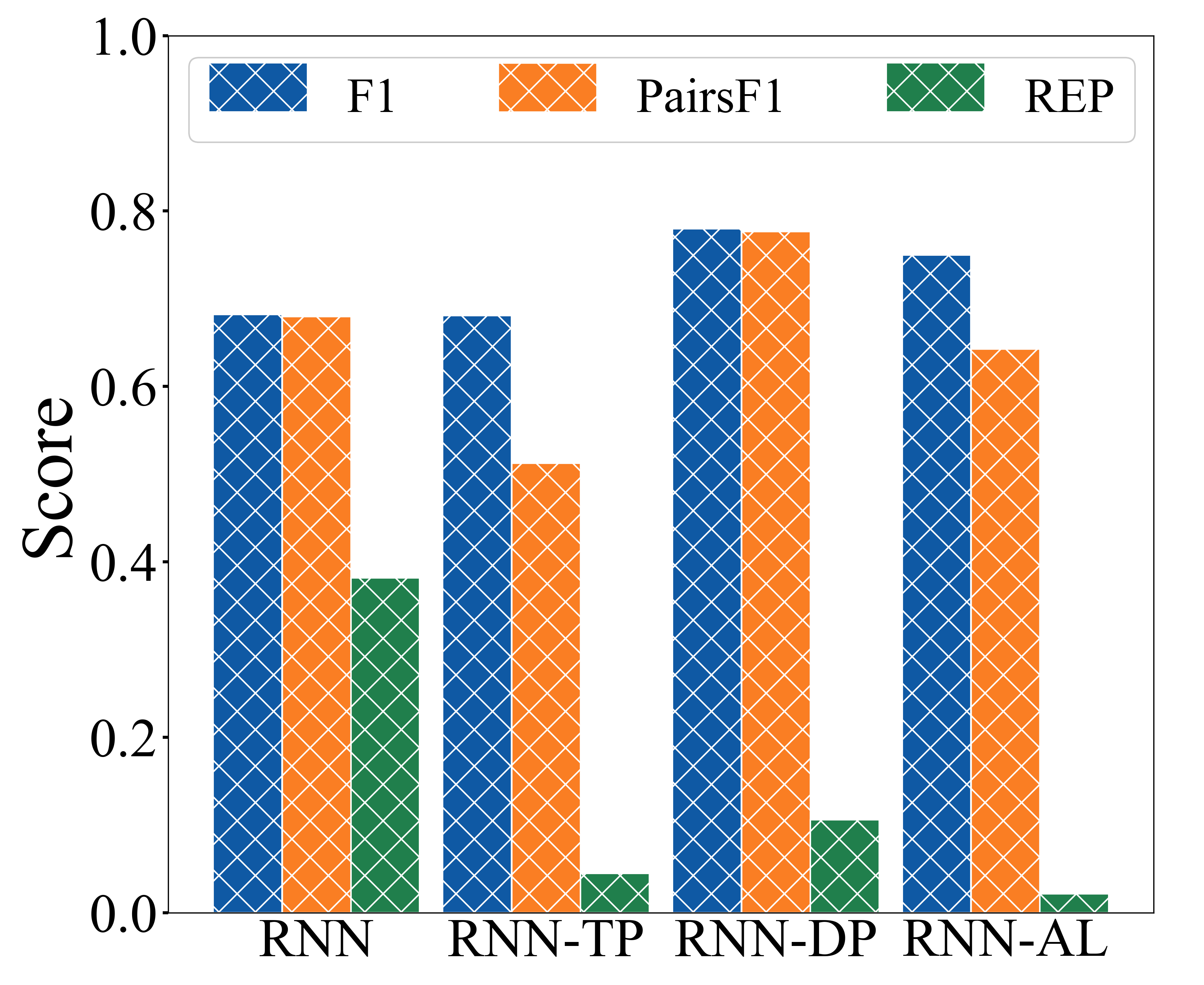}
      \caption{Osaka}
      \label{fig:pics_4}
   \end{subfigure}
   \hfill
   \begin{subfigure}[b]{0.23\textwidth}
      \centering
      \includegraphics[width=\textwidth]{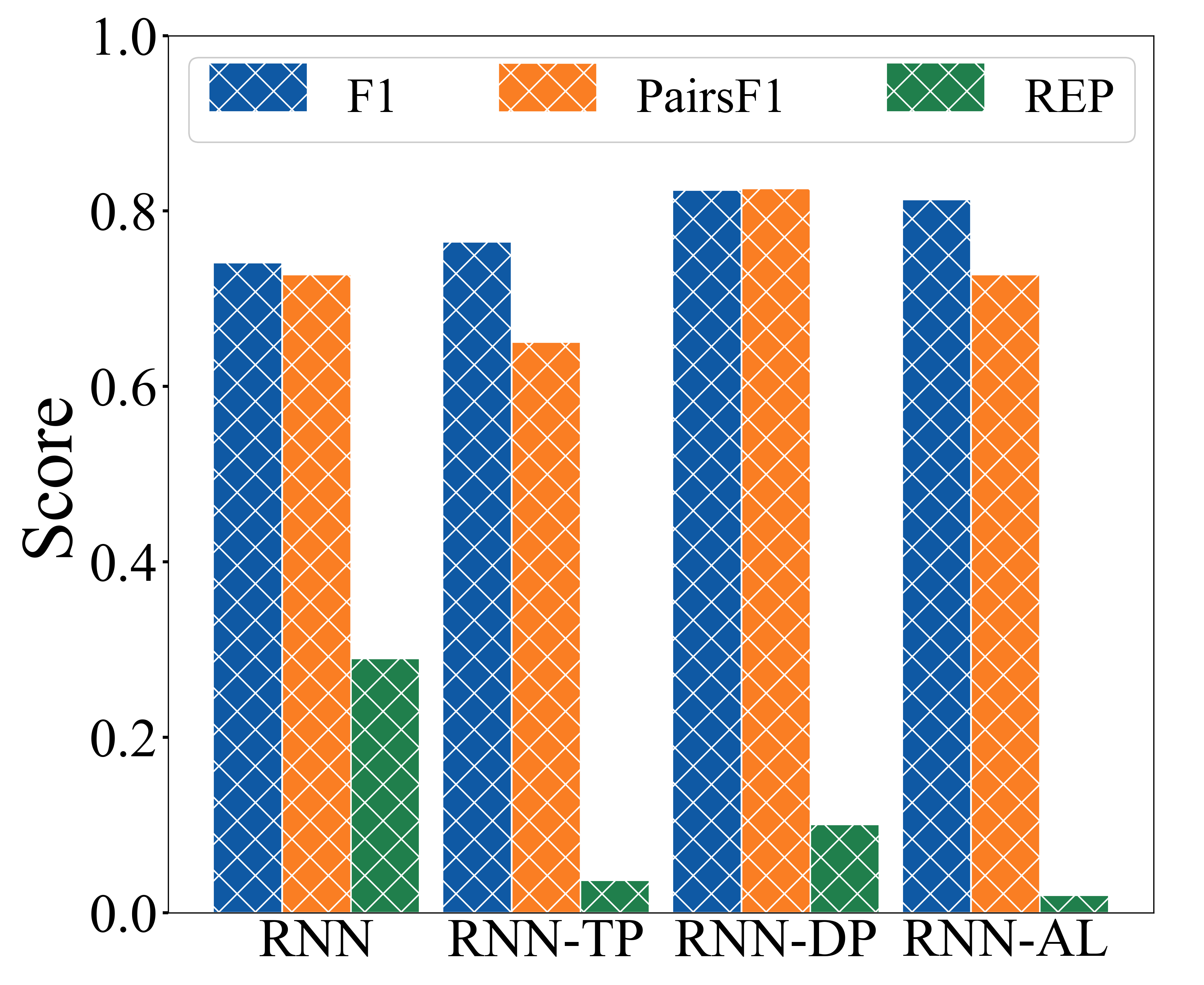}
      \caption{Glasgow}
      \label{fig:pics_5}
   \end{subfigure}
   \caption{The performances in RNN-based model, Note that: (1) RNN (base model with greedy search); (2) RNN-TP (using training penalties, i.e., guiding and drifting); (3) RNN-DP (using a decoding penalty, i.e., adapting); (4) RNN-AL (using all above)}
   \vspace{-5mm}
   \label{fig:pics_4&5}
\end{figure}

\noindent \textbf{Versatility.}
To verify the elements' versatility in Osaka and Glasgow, we designed a traditional RNN-based recommendation architecture following previous works ~\cite{gao2021adversarial,gao2022self}. The results of our approaches are depicted in Figure~\ref{fig:pics_4&5}. We can observe that compared to the basic model of the Transformer, RNN exhibits a higher initial repetition probability. However, with the RNN-AL approach, the repetition is reduced to a level comparable to that of the Transformer, which validates the effectiveness of our proposed method. Additionally, the repetition rate with RNN-TP is lower than that of RNN-DP, aligning with the conclusions drawn from the previous ablation studies.

\section{Concluding Remarks}
In this study, we investigate two causes of TRPs and mitigate them by modifying logits during the training and sampling phases and deploying a cycle-aware predictor. This predictor incorporates guiding, drifting, and adapting mechanisms to resolve these issues. Experimental results effectively demonstrate the efficacy and versatility of the method we propose. Our future work will focus on (1) exploring novel architectures beyond existing frameworks to eliminate TRPs further and (2) designing additional plug-and-play penalty mechanism modules.

\section*{Acknowledgement}
This work was supported by the National Natural Science Foundation of China (Grant No.62176043 and No.62072077) and the Grant SCITLAB-30002 of Intelligent Terminal Key Laboratory of Sichuan Province.

\balance
\bibliographystyle{ACM-Reference-Format}
\bibliography{reference}

\end{document}